\documentclass{PoS}

\newcommand{\p}{\partial}
\newcommand{\pslash}{p\kern-1ex /}
\newcommand{\lslash}{l\kern-1ex /}
\newcommand{\kslash}{k\kern-1ex /}
\newcommand{\dslash}{\p\kern-1.2ex /}
\newcommand{\Dslash}{{\cal D}\kern-1.5ex /}
\newcommand{\Aslash}{A\kern-1.2ex /}

\newcommand{\tr}{{\rm tr}}
\newcommand{\Tr}{{\rm Tr}}

\newcommand{\bea}{\begin{eqnarray}}
\newcommand{\eea}{\end{eqnarray}}
\newcommand{\vol}{\Omega}

\newcommand{\BAN}{\begin{eqnarray*}}
\newcommand{\EAN}{\end{eqnarray*}}
\newcommand{\NTU}{
  Physics Department, Center for Theoretical Sciences, 
  and National Center for Theoretical Sciences, 
  National Taiwan University, Taipei~10617, Taiwan  
}

\newcommand{\RIKEN}{
  Theoretical Physics Laboratory, RIKEN,
  Wako 351-0198, Japan
}

\newcommand{\Tsukuba}{
  Graduate School of Pure and Applied Sciences, University of Tsukuba,
  Tsukuba 305-8571, Japan
}

\newcommand{\BNL}{
  Riken BNL Research Center, Brookhaven National Laboratory, Upton,
  NY11973, USA
}

\newcommand{\KEK}{
  High Energy Accelerator Research Organization (KEK),
  Tsukuba 305-0801, Japan
}

\newcommand{\GUAS}{
  School of High Energy Accelerator Science,
  The Graduate University for Advanced Studies (Sokendai),
  Tsukuba 305-0801, Japan
}

\newcommand{\YITP}{
  Yukawa Institute for Theoretical Physics, 
  Kyoto University, Kyoto 606-8502, Japan
}

\newcommand{\RCAS}{
  Research Center for Applied Sciences, Academia Sinica,
  Taipei~115, Taiwan
}
\title{Topological susceptibility in 2-flavor lattice QCD with fixed topology}

\ShortTitle{Topological susceptibility in 2-flavor lattice QCD with fixed topology}

\author{\speaker{T.W.~Chiu}$^{,a}$\thanks{Email: twchiu@phys.ntu.edu.tw},
        S.~Aoki$^{b,c}$, 
        H.~Fukaya$^d$, 
        S.~Hashimoto$^{e,f}$,
        T.H.~Hsieh$^g$,
        T.~Kaneko$^{e,f}$,
        H.~Matsufuru$^{e}$,
        J.~Noaki$^{e}$,
        K.~Ogawa$^{a}$,
        T.~Onogi$^{h}$,
        N.~Yamada$^{e,f}$ (JLQCD and TWQCD Collaborations)   
        \\
         \llap{$^a$}\NTU      \\
         \llap{$^b$}\Tsukuba  \\
         \llap{$^c$}\BNL      \\
         \llap{$^d$}\RIKEN    \\
         \llap{$^e$}\KEK      \\
         \llap{$^f$}\GUAS      \\
         \llap{$^g$}\RCAS     \\
         \llap{$^h$}\YITP     \\
}

\abstract{

 We determine the topological susceptibility $ \chi_t $ in the trivial
 topological sector generated by lattice simulations of two-flavor
 QCD with overlap Dirac fermion, on a $16^3 \times 32$ lattice with 
 lattice spacing $\sim$ 0.12~fm, at six sea quark masses $m_q$ 
 ranging from $m_s/6$ to $m_s$ (where $m_s$ is the physical strange quark mass).
 The $ \chi_t $ is extracted from the plateau (at large time separation)
 of the time-correlation function of the
 flavor-singlet pseudoscalar meson ($\eta'$),
 which arises from the finite size effect due to fixed topology.
 In the small $m_q$ regime, our result of $\chi_t$ is proportional to $m_q$
 as expected from chiral effective theory. 
 Using the formula $\chi_t=m_q\Sigma/N_f$ by Leutwyler-Smilga, we obtain the
 chiral condensate in $N_f=2$ QCD as 
 $\Sigma^{\overline{\mathrm{MS}}}(\mathrm{2~GeV})
 =[252(5)(10) \mathrm{MeV}]^3 $, in good agreement with
 our previous result obtained in the $\epsilon$-regime. 
}

\FullConference{The XXV International Symposium on Lattice Field Theory\\
                July 30 - August 4 2007\\
                Regensburg, Germany}

\begin{document}

\section{Introduction}

In Quantum Chromodynamics (QCD), the topological susceptibility
($ \chi_t $) is the most crucial quantity to measure the
topological charge fluctuations of the QCD vacuum,
which plays an important role in breaking the $ U_A(1) $ symmetry.
Theoretically, $ \chi_t $ is defined as
\bea
\label{eq:chi_t}
\chi_{t} = \int d^4 x  \left< \rho(x) \rho(0) \right>
\eea
where
\BAN
\rho(x) = \frac{1}{32 \pi^2} \epsilon_{\mu\nu\lambda\sigma}
                             \tr[ F_{\mu\nu}(x) F_{\lambda\sigma}(x) ]
\EAN
is the topological charge density ($\propto$ axial anomaly)
expressed in term of the matrix-valued field tensor $ F_{\mu\nu} $.

With mild assumptions, Witten \cite{Witten:1979vv} and
Veneziano \cite{Veneziano:1979ec}
obtained a relationship between the topological susceptibility
in the quenched approximation and the mass of $ \eta' $ 
meson (flavor singlet) in full QCD with $ N_f $ degenerate flavors, 
namely, $ \chi_t(\mbox{quenched}) = f_\pi^2 m_{\eta'}^2/(4 N_f) $
where $ f_\pi = 131 $ MeV, the decay constant of pion.
This implies that the mass of $ \eta' $ is essentially due to
the axial anomaly relating to non-trivial topological charge
fluctuations, which can turn out to be nonzero even in the chiral limit,
unlike those of the (non-singlet) approximate Goldstone bosons.

Using the Chiral Perturbation Theory (ChPT), 
Leutwyler and Smilga \cite{Leutwyler:1992yt}
obtained the following relation in the chiral limit,
\bea
\label{eq:LS}
\chi_t = \frac{m_q \Sigma}{N_f} + {\cal O}(m_q^2)
\eea
where $ m_q $ is the quark mass, and $ \Sigma $ is the chiral condensate.
In other words, as $ m_q \to 0 $, the topological susceptibility
is suppressed due to internal quark loops.
Most importantly, (\ref{eq:LS}) provides a viable way
to extract $ \Sigma $ from $ \chi_t $ in the chiral limit.

From (\ref{eq:chi_t}), one obtains
\BAN
\chi_t = \frac{\left< Q_t^2 \right>}{\Omega}, \hspace{4mm}
Q_t \equiv  \int d^4 x \rho(x)
\EAN
where $ \Omega $ is the volume of the system, and
$ Q_t $ is the topological charge (which is an integer for QCD).
Thus, one can obtain $ \chi_t $ by counting the number of
gauge configurations for each topological sector.
Obviously, for a set of gauge configurations in the
topologically-trivial sector with $ Q_{t} = 0 $,
it gives $ \chi_t = 0 $.
However, even for a topologically-trivial gauge configuration,
it may possess non-trivial topological excitations in sub-volumes.
Thus, one can investigate whether there are topological excitations
within any sub-volumes, and to measure $ \chi_t $
using the correlation of the topological charges of two sub-volumes 
\cite{Fukaya:2004kp}.

In general, for any topological sector with $ Q_t $,
using saddle point expansion on the QCD partition function
in a finite volume, it can be shown that \cite{Aoki:2007ka}
(see also \cite{Brower:2003yx})
\bea
\label{eq:rho_rho}
\lim_{|x| \to \infty} \left< \rho(x) \rho(0) \right> =
\frac{1}{\vol} \left( \frac{Q_t^2}{\vol} - \chi_t
                     -\frac{c_4}{2 \chi_t \vol} \right)
 + {\cal O}(\vol^{-3})
\eea
where
\BAN
c_4 = -\frac{1}{\vol} \left[   \langle Q_t^4 \rangle_{\theta=0}
                            -3 \langle Q_t^2 \rangle_{\theta=0}^2 \right]
\EAN

Thus, one can consider two spatial sub-volumes at
time slices $ t_1 $ and $ t_2 $, and to measure their
time-correlation function
\BAN
C(t_1 - t_2) = \langle Q(t_1) Q(t_2) \rangle
= \sum_{\vec{x_1},\vec{x_2}} \left<\rho(x_1) \rho(x_2) \right>
\EAN
where the summations run over the spatial volumes at $ t_1 $
and $ t_2 $ respectively. 
Then its plateau at large $ |t_1 - t_2 | $ can be used
to extract $ \chi_t $, provided that $ | c_4 | \ll 2 \chi_t^2 \vol $.

However, for lattice QCD, it is difficult to extract $ \rho(x) $ and $ Q_t $
unambiguously from the gauge link variables, due to their rather
strong fluctuations.

To circumvent this difficulty, one may consider
the Atiyah-Singer index theorem
\cite{Atiyah:1968mp}
\bea
\label{eq:AS_thm}
Q_t = n_+ - n_- = \mbox{index}({\cal D})
\eea
where $ n_\pm $ is the number of zero modes of the massless Dirac
operator $ {\cal D} \equiv \gamma_\mu ( \partial_\mu + i g A_\mu) $
with $ \pm $ chirality. Since $ {\cal D} $ is anti-Hermitian and chirally
symmetric, its nonzero eigenmodes must come in complex conjugate pairs
(i.e., $ {\cal D} \phi = i \lambda \phi $ implies
       $ {\cal D} \gamma_5 \phi = -i \lambda \gamma_5 \phi $, 
for $ \lambda = \lambda^* \ne 0 $)
with zero chirality ($ \int d^4 x \phi^{\dagger} \gamma_5 \phi = 0 $).
Thus one can obtain the identity
\bea
\label{eq:index_Q1}
n_+ - n_- = \int d^4 x \ m~\tr [ \gamma_5 ({\cal D} + m)^{-1}(x,x)]
\eea
by spectral decomposition, where the nonzero modes drop out
due to zero chirality. In view of (\ref{eq:AS_thm}) and (\ref{eq:index_Q1}),
one can regard $ m~\tr [ \gamma_5 ({\cal D} + m)^{-1}(x,x)] $
as topological charge density, to replace $ \rho(x) $
in the measurement of $ \chi_t $.

For lattice QCD, it is well-known that the overlap Dirac operator
\cite{Neuberger:1997fp} in a topological non-trivial gauge background
possesses exact zero modes (with definite chirality) satisfying
the Atiyah-Singer index theorem. Writing the massive 
overlap Dirac operator as
\BAN
D(m) = \left( m_0 + \frac{m}{2} \right) +
\left( m_0 - \frac{m}{2} \right) \gamma_5 \frac{H_w(-m_0)}{\sqrt{H_w^2(-m_0)}}
\EAN
where $ H_w(-m_0) $ is the standard Hermitian Wilson operator with negative
mass $ -m_0 $ ($ 0 < m_0 < 2 $), then
the topological charge density can be defined as
\BAN
\rho_1(x) = m~\tr[ \gamma_5 ( D_c + m)^{-1}_{x,x}]
\EAN
where $ (D_c + m )^{-1} $ is the valence quark propagator with quark
mass $ m $, and $ D_c $ is a chirally symmetric and non-local operator, 
relating to $ D(0) $ by $ D_c = D(0) [1 - D(0)/(2m_0)]^{-1} $ 
\cite{Chiu:1998gp}.
Note that $ (D_c + m)^{-1} $ is exponentially-local
for sufficiently smooth gauge background and nonzero $ m $. 
Here $ \rho_1(x) $ is justified to be topological charge density,
since it can be shown that (see e.g., \cite{Chiu:2002xm})
\bea
\label{eq:Q1_index}
\sum_x \rho_1(x) =  m~\Tr[ \gamma_5 ( D_c + m)^{-1}_{x,x}]
= n_+ - n_-
\eea
which is similar to its counterpart in continuum (\ref{eq:index_Q1}),
where $ \Tr $ denotes trace over Dirac, color and lattice spaces.

\begin{figure}[htb]
\begin{center}
\begin{tabular}{@{}cc@{}}
\includegraphics*[height=6cm,width=7cm]{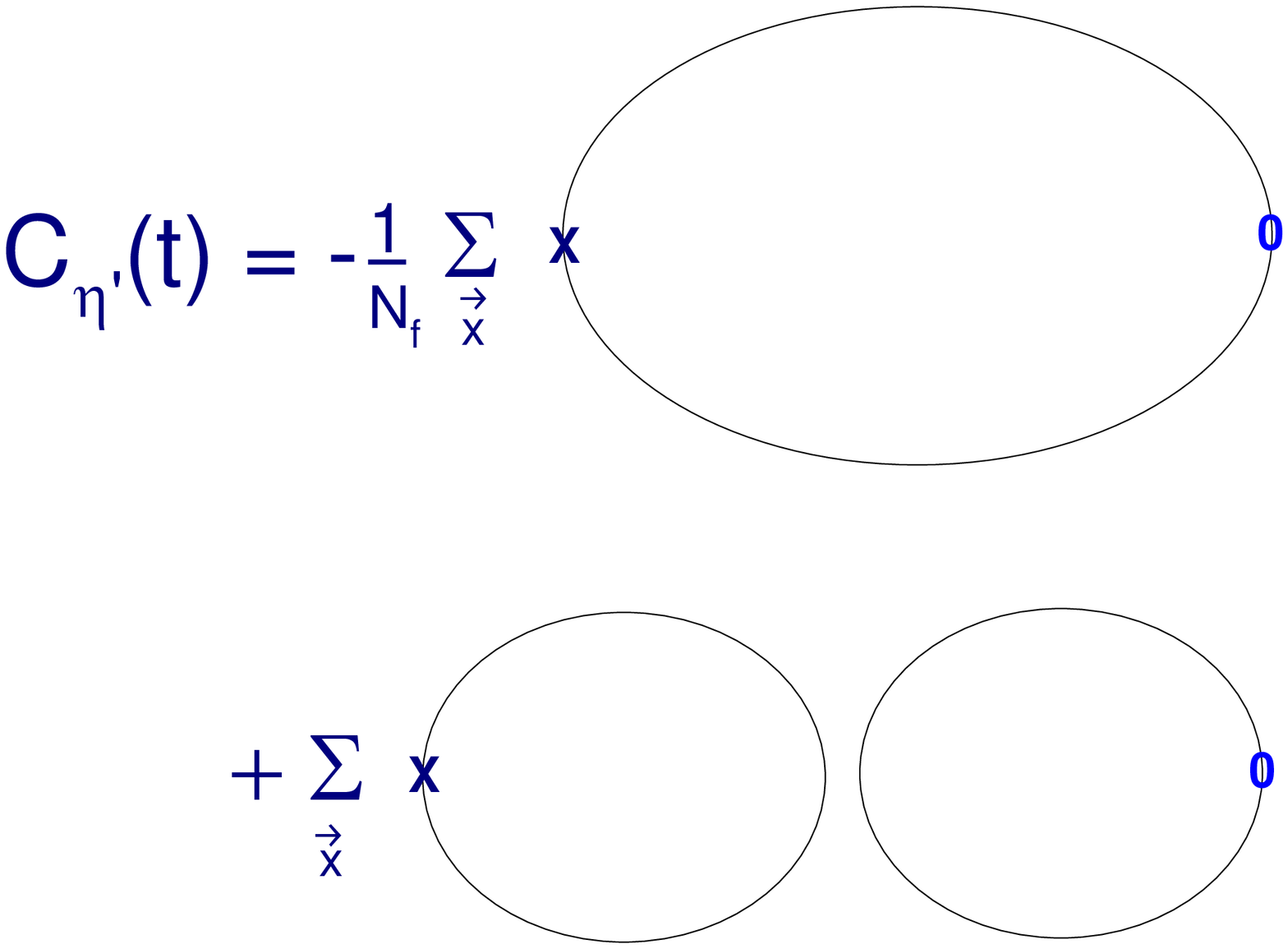}
&
\includegraphics*[height=7cm,clip=true]{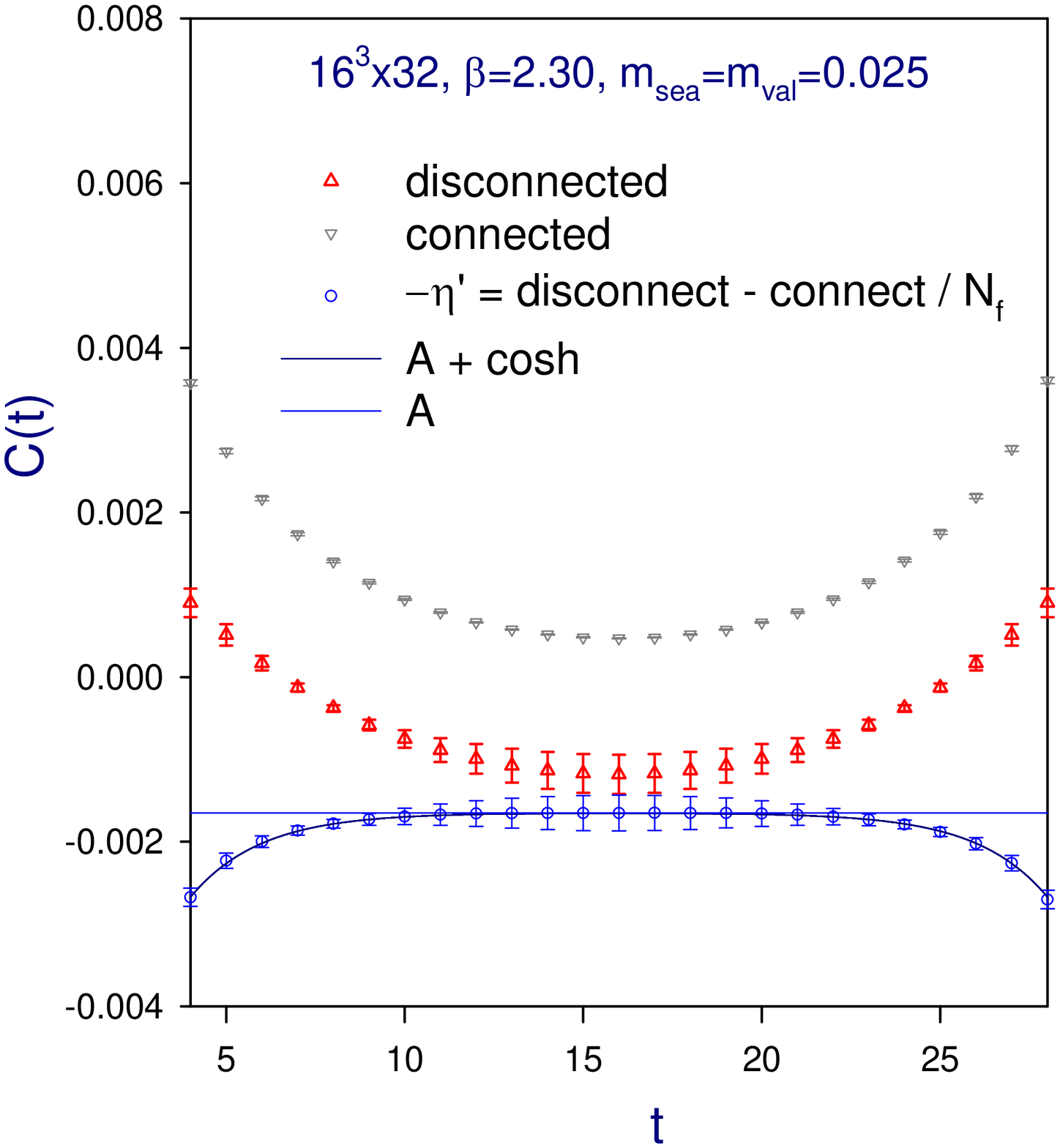}
\\ (a) & (b)
\end{tabular}
\caption{
(a)
   A schematic diagram for the time-correlation function 
   of the flavor singlet pseudoscalar meson operator. 
   Each solid line denotes the valence quark propagator $ (D_c + m_q)^{-1} $.
(b) 
   The time-correlation function of the 
   flavor singlet (circles),  
   and its connected (triangle down) and disconnected (triangle up)
   contributions. Data for $m_q=0.025$ are shown.
}
\label{fig:etap}
\end{center}
\end{figure}

Now we can replace $ \rho(x) $ with $ \rho_1(x) $, and use
(\ref{eq:rho_rho}) to extract $ \chi_t $ for any topological sector.
However, on a finite lattice,
it is contaminated by $ m_\pi $, $ m_{\eta'} $ and any states which
can couple to $ \langle \rho_1(x) \rho_1(0) \rangle $.
A better alternative is to compute the correlator of the flavor-singlet
pseudoscalar meson $ \eta' $, which behaves as \cite{Fukaya:2004kp,Aoki:2007ka}
\bea
\label{eq:etap_etap}
\lim_{|x| \gg 1} m_q^2 \left< \eta'(x) \eta'(0) \right> & \simeq &
\frac{1}{\vol} \left( \frac{Q_t^2}{\vol} - \chi_t -
                      \frac{c_4}{2 \chi_t \vol} \right) 
               +  {\cal O}( e^{-m_{\eta'} |x|})
               + {\cal O}(\vol^{-3})
\eea
Then the time-correlation function of $ \eta' $
(see Fig.~\ref{fig:etap}(a)) is fitted to
$ A + B( e^{-M t} + e^{-M(T-t)} ) $
to extract the constant
$ A = \frac{1}{m_q^2} \frac{1}{T}
    \left( \frac{Q_t^2}{\vol} - \chi_t -
           \frac{c_4}{2 \chi_t \vol} \right) $
and $ \chi_t $, provided that $ |c_4| \ll 2 \chi_t^2 \vol  $.

\section{Lattice Setup}

Simulations are carried out for two-flavor ($N_f=2$) QCD on a $16^3\times 32$
lattice at a lattice spacing $\sim$ 0.12~fm. 
For the gluon part, the Iwasaki action is used at $\beta$ = 2.30, 
together with unphysical Wilson fermions and associated twisted-mass ghosts
\cite{Fukaya:2006vs}.
The unphysical degrees of freedom generate a factor 
$\det[H_w^2(-m_0)/(H_w^2(-m_0)+\mu^2)]$ in the partition function 
(we take $m_0=1.6$ and $\mu=0.2$)
that suppresses the near-zero eigenvalue of $H_w(-m_0)$ and thus
makes the numerical operation with the overlap operator  
substantially faster. 
Furthermore, since the exact zero eigenvalue is forbidden, the global
topological change is preserved during the molecular dynamics evolution of the
gauge field.  Our main runs are performed at $Q=0$, while $Q=-2$ and
$-4$ configurations are also generated at one sea quark
mass in order to check the consistency as described below.

For the sea quark mass $m_q$ we take six values: 0.015, 0.025, 0.035, 0.050,
0.070, and 0.100 that cover the mass range $m_s/6$--$m_s$ with $m_s$ the
physical strange quark mass.
After discarding 500 trajectories for thermalization, we accumulate 
10,000 trajectories in total for each sea quark mass.
In the calculation of $\chi_t$, we take one configuration every 20
trajectories, thus we have 500 configurations for each $ m_q $.
For each configuration, 50 pairs of lowest-lying eigenmodes 
of the overlap-Dirac operator $D(0)$ are calculated using the
implicitly restarted Lanczos algorithm and stored for the later use.

\section{Results}

For the connected diagram (see Fig.~\ref{fig:etap}(a)), the pion correlator 
is computed using the conjugate gradient algorithm with a low-mode
preconditioning. Low-modes are also used for averaging over source points
\cite{DeGrand:2004qw}, which significantly improves the statistical signal.
For the disconnected diagram, the quark propagator is represented by the
eigenmode decomposition and approximated by the 50 conjugate pairs of the 
low-lying eigenmodes.
The quark propagator is then obtained for any source point without extra
computational cost, and the disconnected loops can be calculated with an
average over the source point.
The truncation is motivated by the expectation that the long distance
correlation is dominated by the low-lying fermion modes; its validity has
to be checked numerically (see below).

In Fig.~\ref{fig:etap}(b), we plot $C_{\eta'}(t)$ together with those of
connected and disconnected parts for $m_q = 0.025$.
The curve is a fit to a function $A+B(e^{-Mt}+e^{-M(T-t)})$ with data for
$C_{\eta'}(t)$ in the range $t\in [4,28]$. 
The horizontal line is a fitted constant $A=1.70(13)\times 10^{-3}$, where
the error is estimated using the jackknife method with bin size of
20 configurations, with which the statistical error saturates.
Assuming $|c_4|\ll 2\chi_t^2\vol$, we obtain 
$a^4 \chi_t = 3.40(27) \times 10^{-5}$ at $m_q = 0.025$.

Since the disconnected diagram is computed with only 50 pairs of low-lying 
eigenmodes, we have to check whether they suffice to saturate $C_{\eta'}(t)$.
For the time range $[4,28]$ used for fitting, 
as the number of eigenmodes is increased from 10 to 30, the change of
correlator $|\delta C_{\eta'}|/C_{\eta'}$ is $\sim 3\%$, while from 30 to 50, 
it is only $\sim 0.3\%$, which is less than 8\% of the statistical error.
Thus $C_{\eta'}$ is well saturated with 50 eigenmodes. 
This also holds for all six sea quark masses. 
 
\begin{figure}[tb]
  \centering
  \includegraphics[width=7cm,clip=true]{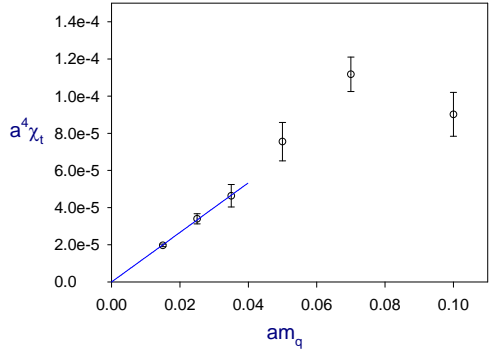}
  \caption{
    Topological susceptibility $ \chi_t a^4 $ 
    versus sea quark mass $m_q a$.}
  \label{fig:chit_mq_Q0}
\end{figure}

In Fig.~\ref{fig:chit_mq_Q0}, we plot the topological susceptibility 
$ \chi_t a^4 $ as a function of the sea quark mass $m_q a $,  
The statistical precision is good enough to find a clear dependence 
on the sea quark mass.
For the smallest three quark masses, 0.015, 0.025, and 0.035, the data are
well fitted by a linear function $F+G m$ with the intercept 
$F = 0.0(1) \times 10^{-5} $ and the slope $G = 0.00133(5)$.
Evidently, the intercept is consistent with zero, in agreement 
with the $\chi$PT expectation (\ref{eq:LS}). 
Equating the slope to $a^3 \Sigma/N_f$, we obtain $a^3\Sigma = 0.0027(1)$.
In order to convert $\Sigma$ to that in the
$\overline{\mathrm{MS}}$ scheme, we calculate the 
renormalization factor $Z_m^{\overline{\mathrm{MS}}}(\mathrm{2~GeV})$
using the non-perturbative renormalization technique
through the RI/MOM scheme \cite{Martinelli:1994ty}.
Our result is 
$Z_m^{\overline{{\mathrm{MS}}}}(\mathrm{2 GeV}) = 0.742(12)$ 
\cite{Noaki:2007es}.
With $ a^{-1} = 1670(20)(20)$ MeV determined with $ r_0 = 0.49 $ fm 
\cite{Kaneko:2006pa}, the value of $ \Sigma $ is transcribed to
$\Sigma^{\overline{{\mathrm{MS}}}}(\mathrm{2 GeV})
  =[252(5)(10) \mathrm{MeV}]^3 $, 
which is in good agreement with our previous result
$ [251(7)(11) \mbox{ MeV}]^3 $ 
\cite{Fukaya:2007fb}
obtained in the $\epsilon$-regime from the low-lying eigenvalues.
The errors represent a combined statistical error
($a^{-1}$ and $Z_m^{\overline{\mathrm{MS}}}$) and
the systematic error estimated from the higher order effects 
({\it e.g.}, $ c_4 $ term), respectively.
Since the calculation is done at a single lattice spacing,
the discretization error cannot be quantified reliably, but
we do not expect much larger error because our lattice
action is free from $O(a)$ discretization effects.
Our results of topological susceptibility  
are listed in Table \ref{tab:chit_r04_mqr0}, for 
six sea quark masses respectively.

\begin{table}[th]
\begin{center}
\begin{tabular}{|c|cccccc|}
\hline
$ m_q r_0 $      & 0.0616(8)  & 0.1016(15)  & 0.1412(17)  & 0.1982(22) 
                 & 0.2742(28) & 0.3852(36)  \\
\hline
$ \chi_t r_0^4 $ & $ 5.59(47) \times 10^{-3} $ 
                 & $ 9.26(1.29) \times 10^{-3} $  
                 & 0.0123(22) & 0.0187(34) 
                 & 0.0263(33) & 0.0199(33)  \\
\hline
\end{tabular}
\end{center}
\caption{\label{tab:chit_r04_mqr0}
The values of topological susceptibility $ \chi_t r_0^4 $ 
extracted in the $ Q = 0 $ sector, for 
six sea quark masses respectively.   
}
\end{table}

In principle, $\chi_t$ in (\ref{eq:rho_rho}) is universal for any 
topological sector.
We check the universality of $ \chi_t $ as follows. 
At sea quark mass $m_q=0.050$, we generate 250 configurations with $Q=-2$
and $-4$ respectively in addition to the main run at $Q=0$.  
Then we extract $\chi_t$ from the time-correlation function of $\eta'$, 
similar to the $Q=0$ case. 
Our results for $a^4 \chi_t$ are: 
$\{ 7.4(1.3), 6.4(2.1), 5.9(1.8) \} \times 10^{-5}$
for $Q=\{0,-2,-4\}$ respectively. 
Evidently, $\chi_t$ extracted from different topological sectors 
are consistent with each other within the statistical error.

Finally, we come to the assumption $|c_4|\ll 2\chi_t^2\vol$
in extracting $ \chi_t $ via (\ref{eq:rho_rho}). 
With the formulas derived in \cite{Aoki:2007ka}, we can obtain 
an estimate of the upper bound of $|c_4|/(2\chi_t^2 \vol)$ by 
measuring the two-point correlator $\langle\rho_1(x_1)\rho_1(x_2)\rangle$ and
the four-point correlator
$\langle\rho_1(x_1)\rho_1(x_2)\rho_1(x_3)\rho_1(x_4)\rangle$. 
Our preliminary result is $ |c_4|/(2 \chi_t^2 \vol) < 0.1 $,  
for all six sea quark masses.
Details of this calculation will be presented elsewhere.
We note that the upper bound $ |c_4|/(2 \chi_t^2 \vol) < 0.1 $  
is also consistent with the ratio 
$ |c_4|/\chi_t\simeq 0.3$ obtained in the quenched approximation  
\cite{Giusti:2007tu}.
 
\section{Conclusion}
  
In this paper, we have determined the topological susceptibility $\chi_t$
in two-flavor QCD from a lattice calculation of two-point correlators at a
fixed global topological charge $ Q_t = 0 $. 
The expected sea quark mass dependence of $\chi_t$ from $\chi$PT is clearly
observed with the good statistical precision we achieved, in contrast to the
previous unquenched lattice calculations.
Our result indicates that the topologically non-trivial excitations 
({\it e.g.}, instanton and anti-instanton pairs) 
are in fact locally active in the QCD vacuum, even when the global topological
charge is kept fixed.
The information of these topological excitations is carried by low-lying
fermion eigenmodes if the exact chiral symmetry is preserved on the lattice.
This work demonstrates that Monte Carlo simulation of lattice QCD with 
fixed topology is a viable approach, to be pursued when the topology change
hardly occurs near the continuum limit even with chirally non-symmetric
lattice fermions.
The artifacts due to the fixed topology in a finite volume can be removed to
obtain the physics results in the $\theta$ vacuum, provided that $\chi_t$ has
been determined in the first place \cite{Aoki:2007ka,Brower:2003yx}, as has
been done in this work. 

  Numerical simulations are performed on Hitachi SR11000 and IBM System Blue
  Gene Solution at High Energy Accelerator Research Organization (KEK) under 
  a support of its Large Scale Simulation Program (No.~07-16), and also 
  in part on NEC SX-8 at YITP (Kyoto U), NEC SX-8 at RCNP (Osaka U), 
  and IBM/HP clusters at NCHC and NTU-CC (Taiwan).
  This work is supported in part by the Grant-in-Aid of the
  Japanese Ministry of Education 
  (No.~13135204, 
       15540251,      
       17740171,       
       18034011,      
       18340075,     
       18740167,      
       18840045,      
       19540286,      
  and  19740160)  
  and the National Science Council of Taiwan 
  (No.~NSC95-2112-M002-005,    
       NSC95-2112-M001-072).     

\end{document}